\documentclass[conference]{IEEEtran}
\IEEEoverridecommandlockouts
\usepackage{cite}
\usepackage{amsmath,amssymb,amsfonts}
\usepackage{algorithmic}
\usepackage{graphicx}
\usepackage{textcomp}
\usepackage{bm}
\usepackage[table]{xcolor}
\definecolor{basecolor}{gray}{0.93}
\usepackage{url}
\usepackage{xcolor}
\usepackage{booktabs}
\usepackage{multirow}
\def\BibTeX{{\rm B\kern-.05em{\sc i\kern-.025em b}\kern-.08em
    T\kern-.1667em\lower.7ex\hbox{E}\kern-.125emX}}
\begin{document}

\title{CompressedVQA-AEV:  Full-Reference and No-Reference Quality Assessment Models for Asymmetric Encoded Videos}
\author{Wei Sun\textsuperscript{1}, Xingwei Liu\textsuperscript{1}, Dandan Zhu\textsuperscript{1}, Xiangyang Zhu\textsuperscript{3}, Weixia Zhang\textsuperscript{2}, Guangtao Zhai\textsuperscript{2}\\
\textsuperscript{1}East China Normal University, \textsuperscript{2}Shanghai Jiao Tong University\\
\textsuperscript{3}Shanghai Artificial Intelligence Laboratory} 

\maketitle

\begin{abstract}
This report presents our solutions to the QoMEX 2026 Grand Challenge on Video Quality Assessment for Asymmetric Encoded Videos, comprising a full-reference (FR) model, CompressedVQA-AEV-FR, and a no-reference (NR) model, CompressedVQA-AEV-NR. The FR approach leverages a Swin-B backbone to extract multi-stage similarity statistics between reference and distorted videos for quality prediction. For the NR setting, our model employs complementary frame-level encoders based on SigLIP2 and Swin-B, followed by temporal mean pooling and cross-fold ensembling to estimate perceptual quality without reference data. Our CompressedVQA-AEV-FR achieves first place in the FR track of QoMEX 2026 Grand Challenge, while CompressedVQA-AEV-NR secures fourth place in the NR track, demonstrating the effectiveness of our proposed models. The code is available at \url{https://github.com/sunwei925/CompressedVQA-AEV}.
\end{abstract}

\section{Introduction}

Video quality assessment (VQA)~\cite{min2024perceptual,sun2022deep,sun2024analysis,cao2026vqathinker,cao2026generalizable,sun2025enhancing} is essential for optimizing modern video encoding and streaming systems. Traditional video quality metrics such as VMAF~\cite{vmaf} and ITU-T P.1204.3~\cite{p1204} have been primarily developed and validated on conventionally (symmetrically) encoded content, where all spatial regions are treated with uniform encoding parameters.

With the advancement of semantic segmentation and visual saliency prediction techniques, Region-of-Interest (ROI) encoding~\cite{wu2024roi} has emerged as a promising strategy that allocates higher bitrates to perceptually important regions while compressing background areas more aggressively. Such asymmetric encoding can achieve significant bitrate savings or improve perceived quality at the same bitrate. However, the reliability of existing VQMs on asymmetrically encoded videos—where spatial quality varies substantially across regions—remains largely underexplored.

To promote research in this direction, the QoMEX 2026 Grand Challenge on Video Quality Assessment for Asymmetric Encoded Videos~\cite{zhu2026qomex_gc_vqa} was organized, providing the Sport-ROI dataset containing both symmetrically and asymmetrically encoded sports videos with subjective quality annotations. Participants were invited to develop full-reference (FR) and no-reference (NR) models that generalize across both encoding paradigms.

In this report, we present our solutions to both tracks. For the FR track, we propose CompressedVQA-AEV-FR, which extracts multi-stage hierarchical features from a Swin Transformer~\cite{liu2021swin} backbone and computes per-stage texture and structure similarity descriptors between reference and distorted frames. For the NR track, we propose CompressedVQA-AEV-NR, which employs complementary SigLIP2~\cite{tschannen2025siglip} and Swin-B~\cite{liu2021swin} encoders with cross-fold ensembling to predict perceptual quality without reference information. Our FR model achieves first place in the FR track, and our NR model secures fourth place in the NR track, demonstrating the effectiveness of the proposed approaches.

\section{Model Architecture}
\label{sec:model_architecture}

\subsubsection{CompressedVQA-AEV-FR}

CompressedVQA-AEV-FR is a full-reference video quality assessment framework designed to predict the perceptual quality of compressed videos by leveraging both reference and distorted content. The framework follows a multi-stage feature similarity paradigm~\cite{sun2025compressedvqa,sun2021deep,ding2020image}: for each pair of reference and distorted frames, deep hierarchical features are independently extracted through a pretrained visual backbone, and per-stage similarity descriptors capturing texture and structure fidelity are computed and concatenated into a unified quality-aware representation. A lightweight regression head maps this representation to a scalar frame-level quality estimate, and the video-level score is obtained via temporal mean pooling over uniformly sampled frames.

\textbf{Multi-Stage Feature Extraction.}
We employ the Swin Transformer Base (Swin-B)~\cite{liu2021swin} as the visual backbone, which computes self-attention within local shifted windows and progressively merges spatial tokens through a hierarchical architecture. This design enables the model to capture multi-scale spatial distortions---from fine-grained compression artifacts such as blocking and ringing to global structural degradations---that are critical for perceptual quality assessment. The backbone is partitioned into four sequential stages, producing feature maps of increasing semantic abstraction with channel dimensions of 128, 256, 512, and 1024, respectively. In addition to these four deep feature stages, the raw de-normalized input image (3 channels) is retained as a stage-zero representation, yielding five feature levels in total with channel dimensions $\{3, 128, 256, 512, 1024\}$.

\textbf{Per-Stage Similarity Computation.}
For each pair of preprocessed reference and distorted frames $z^r$ and $z^d$, the backbone independently extracts hierarchical feature maps $\{f^r_k\}_{k=0}^{4}$ and $\{f^d_k\}_{k=0}^{4}$. At each stage $k$, two complementary similarity descriptors~\cite{ding2020image} are computed to quantify the fidelity between the reference and distorted representations. The first descriptor measures \textit{texture similarity}~\cite{ding2020image} via global mean alignment~:
\begin{equation}
S_{1,k} = \frac{2\,\mu^r_k\,\mu^d_k + c_1}{(\mu^r_k)^2 + (\mu^d_k)^2 + c_1},
\end{equation}
where $\mu^r_k$ and $\mu^d_k$ denote the spatial means of the reference and distorted feature maps at stage $k$, respectively, and $c_1$ is a small constant for numerical stability. The second descriptor captures \textit{structure similarity}~\cite{ding2020image} via variance and covariance statistics:
\begin{equation}
S_{2,k} = \frac{2\,\sigma^{r,d}_k + c_2}{(\sigma^r_k)^2 + (\sigma^d_k)^2 + c_2},
\end{equation}
where $\sigma^r_k$ and $\sigma^d_k$ are the spatial standard deviations of the respective feature maps, $\sigma^{r,d}_k$ denotes the spatial covariance between them, and $c_2$ is a stability constant. Both descriptors are computed independently per channel, producing vectors of dimension $C_k$ at each stage.

\textbf{Quality Regression.}
The per-stage similarity descriptors from all five levels are concatenated into a single quality-aware feature vector:
\begin{equation}
\mathbf{f} = \left[S_{1,0};\, S_{2,0};\, S_{1,1};\, S_{2,1};\, \ldots;\, S_{1,4};\, S_{2,4}\right],
\end{equation}
yielding a representation of dimension $2 \times (3 + 128 + 256 + 512 + 1024) = 3846$. This vector is passed through a two-layer multi-layer perceptron (MLP) with 128 hidden units and no intermediate nonlinearity to produce the frame-level quality score $\hat{q}_t$.

\textbf{Temporal Pooling.}
Given a video, we uniformly sample $T$ frames to form a sparse temporal representation. Each reference--distorted frame pair is independently forwarded through the feature extraction and regression pipeline. The video-level quality score is computed as the arithmetic mean of all per-frame predictions:
\begin{equation}
\hat{Q} = \frac{1}{T}\sum_{t=1}^{T} \hat{q}_t.
\end{equation}

\subsection{CompressedVQA-AEV-NR}
CompressedVQA-AEV-NR is a no-reference video quality assessment framework for predicting the perceptual quality of compressed videos without relying on pristine reference content. The framework follows a frame-level feature extraction and temporal pooling paradigm: individual frames are first independently encoded by a pretrained visual backbone to obtain per-frame quality estimates, which are then aggregated via mean pooling to yield a single video-level quality score.

We investigate two representative visual backbones for frame-level feature extraction, spanning different architectural paradigms within the Transformer family: a Vision Transformer (ViT)~\cite{dosovitskiy2020image} variant and a Shifted Window Transformer~\cite{liu2021swin}.

\textbf{SigLIP2.}
The first architecture builds upon the visual encoder of SigLIP2~\cite{tschannen2025siglip}, a ViT-Base model with a patch size of 16 that operates at a $384\times384$ input resolution. Pretrained with a sigmoid-based contrastive objective on large-scale image-text pairs, the SigLIP2 encoder learns rich semantic and perceptual representations that transfer effectively to downstream quality-aware tasks. It produces a 768-dimensional feature vector per input frame. A lightweight quality regression head, consisting of two cascaded fully connected layers ($768$ and $128$ neurons) without intermediate nonlinearities, is appended to project the feature representation onto an unbounded scalar quality score.

\textbf{Swin.}
The second architecture employs the Swin Transformer Base (Swin-B)~\cite{liu2021swin}, which computes self-attention within local shifted windows and progressively merges spatial tokens through a hierarchical structure. This design enables the model to capture multi-scale spatial distortions—ranging from fine-grained compression artifacts to global structural degradations—that are critical for perceptual quality assessment. The hierarchical feature stages, followed by layer normalization and global average pooling, yield a 1024-dimensional feature vector per frame. A quality regression head of identical design ($1024$ and $128$ neurons) maps the extracted representation to a scalar quality prediction.

The two backbones are complementary in nature: SigLIP2 leverages vision-language pretraining to encode high-level semantic priors, while Swin-B exploits hierarchical locality to model multi-scale spatial distortions. Both architectures share the same regression head design and temporal pooling mechanism, ensuring that performance differences can be solely attributed to the representational capacity of the visual backbone.

\textbf{Temporal Pooling.}
Both architectures share an identical temporal modeling strategy. Given a video, we first uniformly sample $T$ frames to form a sparse temporal representation. Each frame is independently forwarded through the visual backbone and the quality regression head to obtain a per-frame quality estimate. The video-level quality score is then computed as the arithmetic mean of all per-frame predictions:
\begin{equation}
  Q = \frac{1}{T}\sum_{t=1}^{T} q_t,
\end{equation}
where $q_t$ denotes the predicted quality score of the $t$-th sampled frame.

\section{Training Procedure}

\subsection{Training Dataset}
\label{sec:dataset}

We conduct experiments on the QoMEX-GC dataset training set~\cite{zhu2026qomex_gc_vqa}, which comprises video sequences processed under a variety of compression configurations. The dataset is derived from 6 source reference contents (SRCs) spanning 3 distinct scenes, with each scene contributing 2 SRCs. To evaluate generalization, we adopt a content-separated 3-fold cross-validation protocol: in each fold, 4 SRCs are allocated for training and the remaining 2 are reserved for validation to select the best-performing model.

\subsection{Video Preprocessing}
\label{sec:preprocessing}

\textbf{Temporal Sampling.}
To construct a compact temporal representation, we uniformly sample $T\!=\!8$ frames from each video using linearly spaced indices across the total frame count. When a video contains fewer decodable frames than the target number, the last valid frame is replicated to maintain a fixed-length input. This uniform sampling strategy provides balanced coverage of the temporal extent while keeping computational cost constant regardless of video duration.

\textbf{Spatial Transformations.}
Each sampled frame is first resized such that its shorter side equals 420 pixels, preserving the original aspect ratio. During training, a random $384\!\times\!384$ crop is extracted to introduce spatial diversity and mitigate overfitting. During inference, a deterministic center crop of the same size is applied to ensure reproducibility. All frames are normalized using ImageNet statistics.

\subsection{Training Details}
\label{sec:training_details}

All models are optimized using Adam~\cite{kingma2014adam} with an initial learning rate of $1\!\times\!10^{-5}$ and a weight decay of $1\!\times\!10^{-7}$. The learning rate is decayed by a factor of 0.9 every 5 epochs following a step-wise schedule. We adopt the mean squared error (MSE) loss to supervise the predicted quality scores against the ground-truth mean opinion scores (MOS):
\begin{equation}
\mathcal{L} = \frac{1}{N}\sum_{i=1}^{N}\left(\hat{Q}_i - \mathrm{MOS}_i\right)^2,
\end{equation}
where $\hat{Q}_i$ denotes the predicted quality score for the $i$-th video and $N$ is the number of samples in a mini-batch. Training is conducted for 30 epochs with a batch size of 4. The best checkpoint for each fold is selected based on the highest Spearman Rank-Order Correlation Coefficient (SRCC) on the corresponding validation set.

\subsection{Inference and Ensemble}
\label{sec:inference}

At inference time, we employ ensemble strategies for both the FR and NR tracks to improve prediction robustness.

\textbf{FR Track.}
For full-reference quality assessment, we adopt a cross-fold ensemble by selecting the best-performing Swin-B model from each of the three cross-validation folds, yielding an ensemble of three homogeneous models. Each model independently processes the input reference--distorted video pair and produces a scalar quality estimate. The final prediction is obtained by averaging the three individual scores:
\begin{equation}
  \hat{Q}_{\mathrm{FR}} = \frac{1}{M}\sum_{m=1}^{M} Q_{\mathrm{FR}}^{(m)},
\end{equation}
where $Q_{\mathrm{FR}}^{(m)}$ is the prediction of the $m$-th model and $M\!=\!3$ is the ensemble size. By combining models trained on different content-disjoint partitions, this strategy reduces the variance arising from data-dependent biases and yields more stable and generalizable quality predictions.

\textbf{NR Track.}
For no-reference quality assessment, we employ a cross-fold and cross-architecture ensemble. Specifically, we select the best-performing model from each of the three cross-validation folds, yielding an ensemble of three heterogeneous models: a SigLIP2 model from fold~1, a Swin-B model from fold~2, and a SigLIP2 model from fold~3. Each model independently processes the input video and produces a scalar quality estimate. The final prediction is similarly obtained by averaging:
\begin{equation}
  \hat{Q}_{\mathrm{NR}} = \frac{1}{M}\sum_{m=1}^{M} Q_{\mathrm{NR}}^{(m)},
\end{equation}
where $Q_{\mathrm{NR}}^{(m)}$ is the prediction of the $m$-th model and $M\!=\!3$. This strategy leverages two complementary sources of diversity. First, combining models trained on different content-disjoint partitions reduces the variance arising from data-dependent biases. Second, integrating heterogeneous backbones---one grounded in vision--language pretraining and the other in hierarchical local attention---encourages the ensemble to capture both high-level semantic priors and fine-grained spatial distortions, leading to more accurate and generalizable quality predictions.

\section{Experimental Results}
\begin{table}[t]
\caption{Asymmetric condition results (Sport-ROI test set, 40 videos)}
\label{tab:asym_results}
\begin{center}
\resizebox{1\columnwidth}{!}{
\begin{tabular}{lcccc}
\toprule
\textbf{Metric} & \textbf{SRCC $\uparrow$} & \textbf{PLCC $\uparrow$} & \textbf{KRCC $\uparrow$} & \textbf{RMSE $\downarrow$} \\
\midrule
CompressedVQA-AEV-FR & 0.877 & 0.924 & 0.690 & 0.281 \\
\rowcolor{basecolor} VMAF & 0.856 & 0.865 & 0.649 & 0.368 \\
\rowcolor{basecolor} EQM\_NR & 0.840 & 0.879 & 0.638 & 0.350 \\
CompressedVQA-AEV-NR & 0.830 & 0.939 & 0.651 & 0.252 \\
\rowcolor{basecolor} P.1204.3 & 0.642 & 0.668 & 0.472 & 0.546 \\
\rowcolor{basecolor} PSNR-Y & 0.540 & 0.578 & 0.415 & 0.598 \\
\bottomrule
\end{tabular}
}
\end{center}
\end{table}
\subsection{Test Dataset}

\begin{table*}[htbp]
\caption{Symmetric condition results: Sport-ROI (12 videos), internal dataset (281 videos), and weighted average}
\label{tab:sym_results}
\begin{center}
\resizebox{\textwidth}{!}{
\begin{tabular}{l|cccc|cccc|cccc}
\toprule
& \multicolumn{4}{c|}{\textbf{Sport-ROI}} & \multicolumn{4}{c|}{\textbf{Internal}} & \multicolumn{4}{c}{\textbf{Weighted avg.}} \\
\textbf{Metric} & SRCC $\uparrow$ & PLCC $\uparrow$ & KRCC $\uparrow$ & RMSE $\downarrow$ & SRCC $\uparrow$ & PLCC $\uparrow$ & KRCC $\uparrow$ & RMSE $\downarrow$ & SRCC $\uparrow$ & PLCC $\uparrow$ & KRCC $\uparrow$ & RMSE $\downarrow$ \\
\midrule
\rowcolor{basecolor} EQM\_NR & 0.958 & 0.979 & 0.848 & 0.223 & 0.909 & 0.906 & 0.731 & 0.930 & 0.911 & 0.909 & 0.736 & 0.901 \\
\rowcolor{basecolor} VMAF & 0.972 & 0.982 & 0.909 & 0.203 & 0.874 & 0.873 & 0.682 & 1.068 & 0.878 & 0.877 & 0.691 & 1.032 \\
 CompressedVQA-AEV-FR & 0.958 & 0.973 & 0.848 & 0.253 & 0.859 & 0.868 & 0.668 & 1.087 & 0.863 & 0.872 & 0.676 & 1.053 \\
\rowcolor{basecolor} P.1204.3 & 0.923 & 0.963 & 0.818 & 0.289 & 0.827 & 0.827 & 0.626 & 1.229 & 0.831 & 0.833 & 0.634 & 1.190 \\
\rowcolor{basecolor} PSNR-Y & 0.804 & 0.785 & 0.673 & 0.667 & 0.708 & 0.696 & 0.511 & 1.571 & 0.712 & 0.699 & 0.517 & 1.534 \\
 CompressedVQA-AEV-NR & 0.972 & 0.974 & 0.909 & 0.246 & 0.474 & 0.642 & 0.336 & 1.677 & 0.494 & 0.656 & 0.360 & 1.618 \\
\bottomrule
\end{tabular}
}
\end{center}
\end{table*}

\begin{table*}[htbp]
\caption{Overall results.}
\label{tab:overall_results}
\begin{center}
\resizebox{\textwidth}{!}{
\begin{tabular}{l|cccc|cccc|cccc|c}
\toprule
& \multicolumn{4}{c|}{\textbf{Asymmetric}} & \multicolumn{4}{c|}{\textbf{Symmetric}} & \multicolumn{4}{c|}{\textbf{Overall}} & \\
\textbf{Metric} & SRCC $\uparrow$ & PLCC $\uparrow$ & KRCC $\uparrow$ & RMSE $\downarrow$ & SRCC $\uparrow$ & PLCC $\uparrow$ & KRCC $\uparrow$ & RMSE $\downarrow$ & SRCC $\uparrow$ & PLCC $\uparrow$ & KRCC $\uparrow$ & RMSE $\downarrow$ & \textbf{Time (s)} \\
\midrule
\rowcolor{basecolor} EQM\_NR & 0.840 & 0.879 & 0.638 & 0.350 & 0.911 & 0.909 & 0.736 & 0.901 & 0.875 & 0.894 & 0.687 & 0.626 & 5.20 \\
 CompressedVQA-AEV-FR & 0.877 & 0.924 & 0.690 & 0.281 & 0.863 & 0.872 & 0.676 & 1.053 & 0.870 & 0.898 & 0.683 & 0.667 & 18.42 \\
\rowcolor{basecolor} VMAF & 0.856 & 0.865 & 0.649 & 0.368 & 0.878 & 0.877 & 0.691 & 1.032 & 0.867 & 0.871 & 0.670 & 0.700 & 7.15 \\
\rowcolor{basecolor} P.1204.3 & 0.642 & 0.668 & 0.472 & 0.546 & 0.831 & 0.833 & 0.634 & 1.190 & 0.736 & 0.750 & 0.553 & 0.868 & 35.16 \\
 CompressedVQA-AEV-NR & 0.830 & 0.939 & 0.651 & 0.252 & 0.494 & 0.656 & 0.360 & 1.618 & 0.662 & 0.797 & 0.505 & 0.935 & 25.83 \\
\rowcolor{basecolor} PSNR-Y & 0.540 & 0.578 & 0.415 & 0.598 & 0.712 & 0.699 & 0.517 & 1.534 & 0.626 & 0.639 & 0.466 & 1.066 & -- \\
\bottomrule
\end{tabular}
}
\end{center}
\end{table*}

The proposed models are validated on the test set of the QoMEX 2026 Grand Challenge on Video Quality Assessment for Asymmetric Encoded Videos~\cite{zhu2026qomex_gc_vqa}. The test set includes the Sport-ROI test set, which contains $52$ videos from $2$ held-out SRCs, comprising $40$ asymmetrically encoded videos and $12$ symmetrically encoded videos, and an internal symmetric dataset containing $281$ videos from $20$ SRCs. The models are evaluated on two symmetric sets and one asymmetric set.

\subsection{Evaluation Criteria}

SRCC serves as the primary ranking criterion. Pearson Linear Correlation Coefficient (PLCC), Kendall Rank Correlation Coefficient (KRCC), and Root Mean Square Error (RMSE) are also reported for comparison. PLCC and RMSE are computed after a four-parameter nonlinear mapping~\cite{seshadrinathan2010study}. The official final ranking score is
\begin{equation}
\mathrm{SRCC}_{\mathrm{final}} = 0.5 \, \mathrm{SRCC}_{\mathrm{asym}} + 0.5 \, \mathrm{SRCC}_{\mathrm{sym}},
\label{eq:final_rank}
\end{equation}
where the symmetric term is itself a weighted average:
\begin{equation}
\mathrm{SRCC}_{\mathrm{sym}} =
\frac{12}{293} \, \mathrm{SRCC}_{\mathrm{sym}}^{\mathrm{Sport\mbox{-}ROI}} +
\frac{281}{293} \, \mathrm{SRCC}_{\mathrm{sym}}^{\mathrm{internal}}.
\label{eq:sym_rank}
\end{equation}

\subsection{Baseline Methods}

We compare our proposed models against the following baselines. VMAF~\cite{vmaf} is a full-reference video quality metric developed by Netflix that fuses multiple elementary features (multi-scale VIF, Detail Loss Metric, and motion information) via a support vector machine regressor trained on subjective scores. ITU-T P.1204.3~\cite{p1204} is a standardized no-reference bitstream-based model that predicts perceptual quality from codec-level metadata such as quantization parameters, frame types, and resolution without pixel-level decoding. EQM NR~\cite{chen2024eqm} is a no-reference model that combines encoding parameters with motion activity statistics to estimate compression-induced quality degradation. PSNR-Y computes peak signal-to-noise ratio on the luminance channel, serving as a traditional signal-level fidelity baseline.

\subsection{Experimental Analysis}
We report the performance of baseline and proposed methods on the asymmetric condition dataset in Table~\ref{tab:asym_results}, on the symmetric condition datasets in Table~\ref{tab:sym_results}, and the overall performance in Table~\ref{tab:overall_results}~\footnote{The results are provided by the report of QoMEX 2026 Grand Challenge on Video Quality Assessment for Asymmetric Encoded Videos~\cite{zhu2026qomex_gc_vqa}.}.

\noindent\textbf{Results on Asymmetric Encoding.}
CompressedVQA-AEV-FR achieves the highest correlation on asymmetrically encoded videos, outperforming VMAF and all other baselines. This demonstrates that our multi-stage similarity framework effectively captures spatially non-uniform distortion introduced by ROI-based encoding. CompressedVQA-AEV-NR also performs competitively, achieving the lowest RMSE among all methods.

\noindent\textbf{Results on Symmetric Encoding.}
CompressedVQA-AEV-FR remains competitive with VMAF on symmetric content, while CompressedVQA-AEV-NR exhibits notable performance degradation. This generalization gap likely stems from limited training diversity (only 6 SRCs from 3 sport scenes). The FR model is less susceptible because per-stage similarity computation inherently normalizes content-dependent variations.

\noindent\textbf{Overall Performance.}
CompressedVQA-AEV-FR secures first place in the FR track, confirming that learning-based multi-stage similarity can outperform handcrafted feature fusion even with limited training data. CompressedVQA-AEV-NR ranks fourth among NR submissions, with the gap primarily attributable to limited generalization on symmetric content. The contrasting behavior highlights a fundamental challenge: NR models must implicitly learn both content representation and quality mapping from limited annotations, whereas the FR model circumvents this through content-agnostic similarity features. Future work should explore pretraining on larger-scale datasets to address this limitation.
\section{Conclusion}
In this report, we present CompressedVQA-AEV, comprising a full-reference model based on multi-stage similarity learning and a no-reference model leveraging deep feature regression. CompressedVQA-AEV-FR achieves first place in the FR track of the QoMEX 2025 Grand Challenge, while CompressedVQA-AEV-NR secures fourth place in the NR track, demonstrating the effectiveness of our proposed approaches.

\bibliographystyle{IEEEbib}
\bibliography{icme2025references}

\vspace{12pt}

\end{document}